\documentclass[aps,pra,twocolumn,groupedaddress,floatfix]{revtex4}

\usepackage{graphicx}
\usepackage{textcomp}

\begin{document}

\title{~\vspace{1.5cm}\\ Optical trapping of absorbing particles}


\author{H. Rubinsztein-Dunlop}

\author{T. A. Nieminen}

\author{M. E. J. Friese}
\author{N. R. Heckenberg}

\affiliation{Department of Physics,
The University of Queensland, Brisbane QLD 4072, Australia}


\date{21st November 1996}

\begin{abstract}
\vspace{-5cm}
\noindent
\hspace{-1.5cm}\textbf{Preprint of:}

\noindent
\hspace{-1.5cm}H. Rubinsztein-Dunlop, T. A. Nieminen, M. E. J. Friese,
and N. R. Heckenberg,

\noindent
\hspace{-1.5cm}``Optical trapping of absorbing particles'',

\noindent
\hspace{-1.5cm}\textit{Advances in Quantum Chemistry} \textbf{30},
469--492 (1998)

\hrulefill

\vspace{3cm}

Radiation pressure forces in a focussed laser beam can be used to trap
microscopic absorbing particles against a substrate.  Calculations based
on momentum transfer considerations show that stable trapping occurs before
the beam waist, and that trapping is more effective with doughnut beams.
Such doughnut beams can transfer angular momentum leading to rotation of
the trapped particles.  Energy is also transferred, which can result in
heating of the particles to temperatures above the boiling point of the
surrounding medium.
\end{abstract}


\maketitle

\section{Introduction}

The availability of lasers has enabled the observation of forces due to
light interacting with microscopic objects. In 1970, Ashkin~\cite{ref1}
reported optical trapping of micrometre sized spheres using two opposing
laser beams, and by 1980~\cite{ref2} had proposed many experiments using
focused laser beams, and discussed widely varying possible applications,
including automatic force measurement, particle size measurement using
surface wave resonances in scattering particles, a modified Millikan
experiment, measurement of radiometric forces, and separation and
manipulation of biological particles. In 1986~\cite{ref3} the single-beam
gradient optical trap, or \emph{optical tweezers} was first demonstrated.
Optical tweezers can be used for three-dimensional manipulation of
transparent particles around 1--100\,{\textmu}m in diameter, or other
small particles which behave as reactive dipoles.

The single-beam gradient optical trap consists of a single laser beam,
tightly focused to create a very strong field gradient both radially and
axially, which acts on polarisable particles to cause a dipole force.
Polarisable particles are attracted to the strongest part of the field,
at the beam focus, due to the gradient force.  A scattering force results
from momentum transfer to the particle when light is scattered by it.
Under the right conditions the gradient force can balance the scattering
and gravity forces, to trap particles three-dimensionally in the laser beam.
If the laser beam is not tightly focused, the axial component of the gradient
force will be weak, and only radial trapping will be possible~\cite{ref1,ref4}.

A practical optical tweezers setup usually consists of a laser beam with a
power of a few hundred mW, introduced into a microscope and focused using
a high numerical aperture $100\times$ objective lens.  Both conventional
upright and inverted microscopes are used, the inverted microscope providing
stronger axial trapping due to the scattering force opposing gravity.

The ability to manipulate transparent particles of size 1--100\,{\textmu}m
in a closed sterile environment has been exploited by researchers in the
biological sciences, to move~\cite{ref5}, isolate~\cite{ref6,ref7}, cut
(using UV beams, where biological specimens are highly absorbing)~\cite{ref8},
and perform surgery on cells~\cite{ref9} and to do various forms of analysis.
The very high intensity region at the beam focus can be used for two photon
spectroscopy, which enables analysis of very thin sections of a sample, since
the high intensity region is extremely localised~\cite{ref10}. Using
calibrated forces from optical tweezers, it is also possible to measure
physical properties of specimens, such as the compliance of bacterial
flagella~\cite{ref11}, mechanical properties of a single protein
motor~\cite{ref12} and tube-like motion of a single polymer chain~\cite{ref13}.

Although the particles normally trapped with optical tweezers are highly
transparent, there is usually some absorption occurring.  Studies have been
made of the wavelength dependence of the heating of specimens and the ability
of live specimens to remain viable after having been
trapped~\cite{ref14,ref15,ref16}.

Svoboda and Block~\cite{ref17} reported that small metallic Rayleigh particles
with sizes on the order of tens of nm can be trapped using dipole forces.
However when the absorptivity of micron-sized particles becomes high enough,
radiation pressure becomes much greater than the gradient force, and they can
no longer be trapped using dipole forces.  These absorbing particles are then
affected primarily by radiation pressure, whereby the momentum of the absorbed
light is transferred to the particle.  The momentum of the light field is
normal to the wavefronts, so if the curvature of the wavefront is such that
the resulting force can be resolved into a radially inward force and a force
in the direction of beam propagation, then particles constrained in the
direction of beam propagation (e.g. by a glass microscope slide as in
Figure 1) can be trapped two-dimensionally.

\begin{figure}[ht]
\includegraphics[width=\columnwidth]{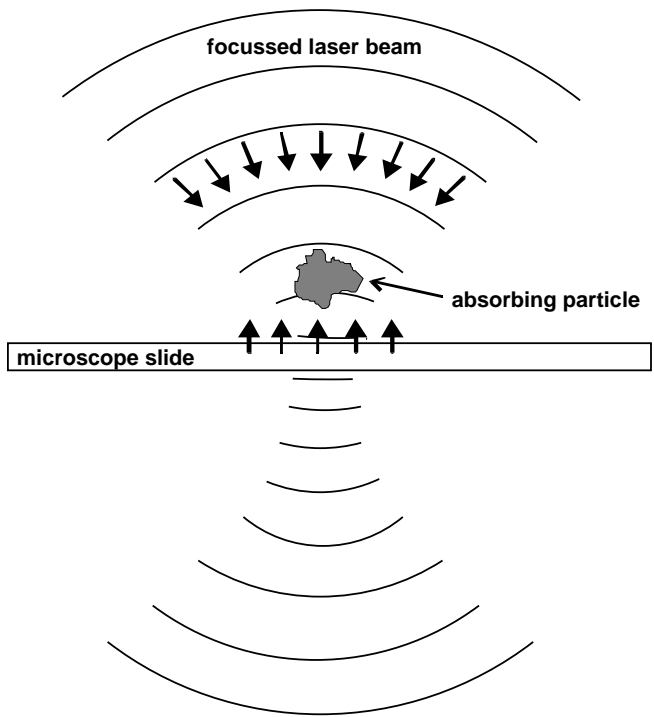}
\caption{Schematic diagram of an absorbing particle in a focussed laser beam.
The curvature of the wavefront is such that the resulting force can be
resolved into a radially inward force and a force in the direction of the
beam propagation.  The particles are constrained in the vertical direction
by the glass microscope slide, and hence, trapped.}
\end{figure}

Trapping of micron-sized reflective particles has been modelled in a similar 
way~\cite{ref18}. On the basis of ray optics, a totally reflective particle
can be shown to be radially constrained when located before the beam focus.
However, the forces produced by reflection are dependent on the angle at
which the laser light impinges on the particle.  Trapping of absorbing
particles is much less dependent on particle geometry as the radiation
force is perpendicular to the wavefront and independent of the orientation
of the particle surface.  In experiments on the trapping of microscopic
reflective particles, heating effects such as bubble formation and
radiometric forces were observed~\cite{ref18}, indicating a noticeable
absorption effect, so absorption of linear momentum also contributes to
the trapping forces acting on these particles.

Using such transfer of linear momentum, absorbing particles such as zinc
dust~\cite{ref19}, CuO particles~\cite{ref20} and ceramic
powder~\cite{ref21,ref22} have been trapped and manipulated. The laser
beam used in most of these experiments was equivalent to a Gauss--Laguerre
LG$_{03}$ mode, a \emph{doughnut} beam.
When a doughnut beam is used for trapping,
the laser intensity is concentrated in a ring of light: this means that the
radiation force along the beam axis is less than for a TEM$_{00}$ mode, and the
radiation trapping is comparatively stronger.

As well as linear momentum, angular momentum is transferred from light to
absorbing particles during trapping. Both \emph{orbital} angular momentum
due to the helical wavefront of a LG$_{03}$ mode~\cite{ref19,ref21,ref23}
and \emph{spin} angular momentum due to the polarisation of the
light~\cite{ref19,ref23,ref24} have been observed to set particles into
rotation.

In this paper we review the work performed on absorbing particles.  Both
Gaussian and Laguerre--Gauss modes  are commonly used.  We outline a model
for the trapping of absorbing particles for these beam types based on the
effects of transfer of momentum from the beam to the particle.  We also
consider in some detail the heating effects.

\section{Laser tweezers experiments}

\begin{figure}[h]
\includegraphics[width=\columnwidth]{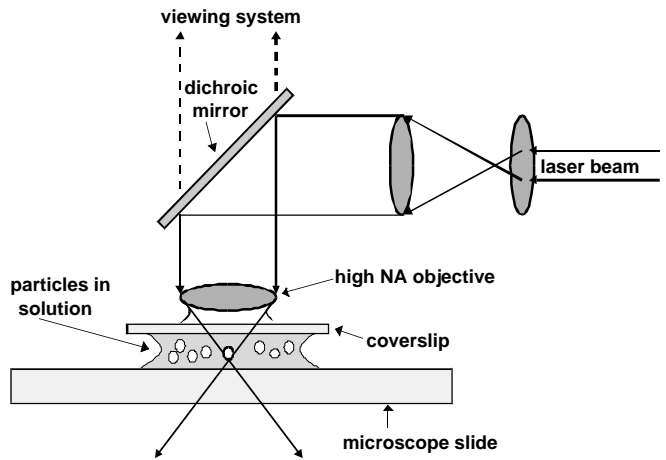}
\caption{Basic experimental setup for laser tweezers.
A dichroic mirror is used to direct the beam into the objective.
The particles to be trapped are suspended in a solution, between a
glass microscope slide and a coverslip. An oil-immersion objective is
used for low distortion and tight focussing of the beam. The laser beam
is expanded to fill the objective in order to produce the smallest spot
and largest intensity gradient to ensure optimal trapping.}
\end{figure}

The most commonly used experimental set up (depicted in Figure 2) for the
single beam optical trap is optimised for trapping transparent particles
and consists of a laser light source directed into a port of an optical
microscope~\cite{ref25}.  A variety of lasers can be used for trapping,
ranging from He-Ne lasers, CW NdYAG lasers, Ti Sapphire, Ar ion lasers to
diode laser sources~\cite{ref26,ref27,ref28,ref29,ref30,ref31,ref32,ref33}.
A single mode laser beam is introduced into the microscope in such a way as
not to interfere with normal microscope function.  It is brought to a tight
focus at (or near) the specimen plane, usually with a high numerical aperture
($\textrm{NA} \ge 1$) oil immersion $100\times$ objective lens.  As
microscope optics are designed to minimise aberrations near the specimen
plane, arranging the optics in such a way that the trap is parfocal with
the specimen allows trapped objects to be visualised and improves the
quality of the trap.  High numerical aperture is essential to maximise
the light intensity gradient near the focal plane and ensure stable
trapping in the axial direction.  The diameter of the laser beam is
normally expanded just before being introduced into the microscope using
a set of appropriate lenses.  This is done in order to ensure exactly
filling, or somewhat over-filling the back pupil of the objective lens.
In this way minimum focal spot size and maximum intensity gradient, and
hence the strongest trapping force on an object to be trapped and
manipulated is achieved.  The expanded beam is deflected to the objective
by a dichroic mirror.  The spot size of the laser beam is of the order of
1--2\,{\textmu}m at the focal plane of the microscope.

Particles to be trapped are placed between a microscope slide and coverslip.
Depending on the type of particle, the solution in which they are kept varies
(water, kerosene etc).  The optical trapping is observed using a CCD camera
mounted on the microscope.

In most experiments the trap has to be moved with respect to the specimen.
This can be done by either moving the specimen or by moving the
beam~\cite{ref34}.  A specimen can be positioned in the $xy$-plane by moving
the microscope stage in the conventional way, leaving the trap fixed on the
optic axis.  When small precise displacements are needed, the sample can be
mounted on an $xy$ piezoelectric stage which is computer controlled.
Movement of the trap in the $z$-direction (depth) is achieved by focusing
the microscope up or down, taking advantage of the parfocality of the trap
and specimen. The $z$-direction movement can also be made in extra fine
steps if either the sample or the objective is placed on a vertical
piezoelectric element.  A change in the vertical position of the trap
with respect to the specimen plane can be achieved by moving an external
lens which controls the beam divergence.  A large number of experiments
involving the trapping of transparent objects with refractive indices higher
than the surrounding medium have been performed using the above described
set-up [for reviews, see~\cite{ref35,ref36,ref37,ref38}].

For transparent (non-absorbing) objects, the strongest trapping is achieved
at the beam waist which is near the object plane of the microscope lens and
therefore trapped particles are in focus when viewed.  The forces acting on
the trapped particle have been studied by calibrating against viscous drag
exerted by fluid flow using escape-force
methods~\cite{ref25,ref26,ref29,ref39,ref40,ref41}.  The force can also be
measured as a function of displacement from the trap centre.  In this way,
the trap stiffness, when the particle in a single-beam gradient optical trap
is modelled as a mass in a 3-D (ellipsoidal) harmonic potential
well~\cite{ref42,ref43,ref44}, can be determined.

\subsection{Absorbing particles}

It was considered for some time, that a strongly absorbing particle with a
high complex index of refraction should be impossible to trap using a
Gaussian beam.  Such a particle would be pushed out of the beam. Therefore,
in a number of experiments hollow beams of different types have been used
to achieve trapping of absorbing or reflecting particles.  In one of the
experiments the trapping was achieved by scanning a beam in a circle using
galvanometer mirrors and in this way producing a hollow beam---an
intensity minimum surrounded by bright circle.  Successful trapping was
obtained with a particle being confined in the intensity minimum region
of the beam~\cite{ref45}.  It has been also demonstrated that optical
levitation of metal particles could be obtained using a TEM$_{01}^\ast$ mode
laser beam~\cite{ref4}.

Recently, we have shown in a series of experiments that two-dimensional
optical trapping of highly absorbing particles can be achieved using a
Gauss--Laguerre LG$_{03}$ mode laser beam~\cite{ref20,ref21,ref22,ref44}.
In these experiments, particles were trapped against a microscope slide in
the converging beam before the waist where the radial component of momentum
at any position is directed radially inwards.  

\subsection{Doughnut beams}

Beams of this type contain a phase singularity which is defined as a point
in an optical field around which the phase of the field changes by an
integer multiple of $2\pi$.  The integer is denoted by $l$ and called the
topological \emph{charge} of the phase singularity.  The charge $l$ is the
\emph{azimuthal mode index} of a Gauss--Laguerre beam.  At the singularity
the phase is undefined and it appears as a dark spot on a bright background.
Beams containing phase singularities can be produced in a variety of ways
such as by transformation of Hermite--Gaussian modes, or using cooperative
mode locking of a laser.  We have shown earlier that beams containing phase
singularities can be conveniently produced using computer generated holograms
with high efficiency~\cite{ref22}.  A phase singularity hologram is similar
to a grating except that it has a defect in the centre of the pattern.
Using computer graphic techniques, a blazed interference pattern with a
central defect can be produced.  This is followed by photoreduction of the
pattern onto a film which results in an amplitude hologram.  The
photoreduced patterns are contact printed onto a holographic plate and the
developed plate is bleached to produce a phase hologram of high efficiency.
In this way we can produce holograms with different order phase singularities.
Angular momentum is associated with the helical structure of the wave
surrounding a singularity so that a linearly polarised beam with a charge
$l$ singularity will carry angular momentum $\hbar l$ per photon.

\subsection{Transfer of angular momentum}

In our experiments using an optical tweezers set-up, only slightly modified
compared to the conventional system described above,  we have shown that
using phase singular fields we can not only trap absorbing particles but
also set them into rotation.  This demonstrates the transfer of the
angular momentum from the light beam to the particles.

The purpose of our experiments was to unambiguously demonstrate transfer
of angular momentum, evaluate the resulting rotation speed of the particles
and investigate the relationship of the angular momentum associated with the
helical structure of the beam to that associated with circular polarisation.

The transfer of angular momentum from light to an absorbing particle can be
understood by considering that the light torque arises from the angular
momentum carried by photons.  The Gauss--Laguerre modes are eigenmodes of
the angular momentum operator $L_z$ and as such have an \emph{orbital}
angular momentum $\hbar l$ per photon.  If the Gauss--Laguerre mode is
circularly polarised we can then also assign it \emph{spin} angular momentum
of $\sigma_z\hbar$ per photon, where $\sigma_z$ is $\pm 1$ (or zero for
linear polarisation).  So, in a paraxial approximation, considering the
total number of photons absorbed per second, the torque due to both the
polarisation and the helical Poynting vector of the Gauss--Laguerre mode
is given by
\begin{equation}
\Gamma = \frac{P_\mathrm{abs}}{\omega}(l+\sigma_z)
\end{equation}
where $P_\mathrm{abs}$ is the power absorbed by the particle and $\omega$
is the frequency of light.  Barnett and Allen~\cite{ref46} have developed
a general nonparaxial theory according to which the torque will be given by
\begin{equation}
\Gamma = \frac{P_\mathrm{abs}}{\omega}
\left\{ (l+\sigma_z) + \sigma_z \frac{2p+l+1}{2kz_r} \right\}
\end{equation}
where $k$ is the wave number, $p$ and $l$ are the radial and azimuthal
mode indices and $z_r$ is a length term, which in the paraxial limit is
associated with the Rayleigh range.  However it can be shown that the
mixed term in the above equation becomes significant only when the beam
is very strongly focused.  Even in the experimental situation, when the
Gauss--Laguerre LG$_{03}$ mode is used and is focused to approximately
2\,{\textmu}m size waist, the contribution of this term can be neglected.

The experimental set-up we used for studies of trapping of absorbing
particles is similar to the one described in the earlier part of this
paper with only a few modifications: a Gaussian TEM$_{00}$ beam from a
17\,mW He--Ne laser is passed through a computer generated phase hologram,
which produces a Gauss--Laguerre LG$_{03}$ mode in the far field.  This
plane polarised, helical doughnut beam is then introduced in the usual
way into the microscope (see Figure 3).

\begin{figure}[ht]
\includegraphics[width=\columnwidth]{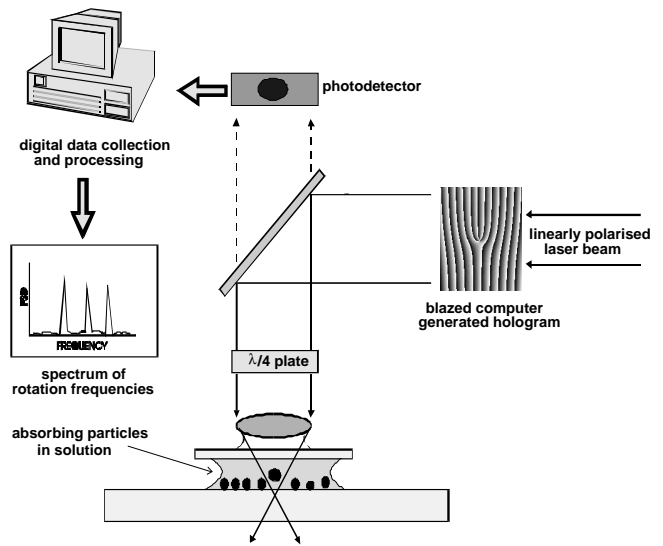}
\caption{Experimental arrangement for measurement of rotation frequency
of absorbing particles.
Scattered light from rotating particles is measured with a photodiode.
The photodiode signal is digitally sampled, then this data is processed
using Fourier analysis to produce spectra of rotation frequencies.}
\end{figure}

The trapping of absorbing particles was performed using irregular black
ceramic particles and CuO particles dispersed in kerosene.  We have also
trapped slightly absorbing latex spheres which were in clumps so that their
rotation could be more easily observed.

We found that most efficient trapping was achieved with the particles
slightly above the focal plane of the doughnut beam.  A variety of
experiments was performed.  Firstly, trapping and rotation could be
observed using a CCD camera fitted to the microscope.  In this set of
studies the absorbing particles were trapped when the hologram of charge
3 was placed in the beam.  A particle was trapped and observed to rotate.
The rotation could be observed for a long period of time.  Subsequently
the hologram was moved sideways so that the particle was illuminated by
a Gaussian beam and stopped rotating.  As the hologram used here is blazed
the sign of the doughnut can be simply reversed by turning the hologram
around.  In principle, on the reversal of the hologram, the particle should
rotate in the opposite direction.  However, in the process of turning the
hologram, the particle is no longer trapped and so on the reversal of the
hologram, it is not certain that the retrapped particle is the same as
before.  It is, however, certain that on the reversal of the hologram,
all of the trapped particles rotate in the opposite direction.  Another
method of reversing the helicity of the doughnut beam is to introduce a
Dove prism into the beam path between the hologram and the microscope.
As the beam undergoes one reflection in the prism, its helicity is
reversed and the prism can be adjusted in such a way so that the beam
will be undeviated.  With the Dove prism in the beam path, we now trap
the particle, detect its direction of rotation and then quickly remove
the Dove prism so that the same particle is still trapped.  The direction
of rotation is reversed~\cite{ref19,ref47}. The sequence of events is
recorded and reproduced frame by frame (see Figure 4).

\begin{figure}[ht]
\begin{tabular}{cccc}
\includegraphics[width=0.22\columnwidth]{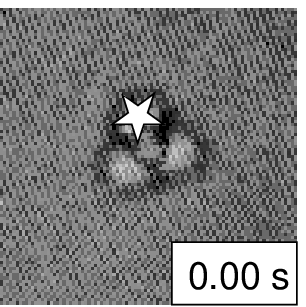}&
\includegraphics[width=0.22\columnwidth]{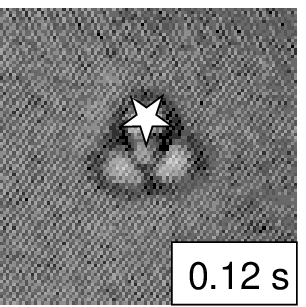}&
\includegraphics[width=0.22\columnwidth]{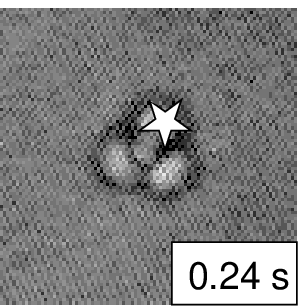}&
\includegraphics[width=0.22\columnwidth]{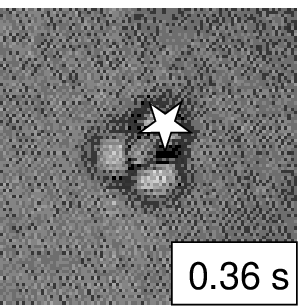}\\
\includegraphics[width=0.22\columnwidth]{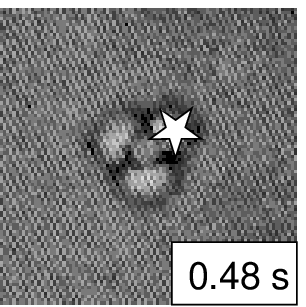}&
\includegraphics[width=0.22\columnwidth]{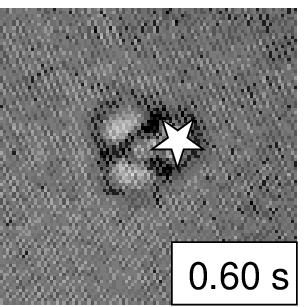}&
\includegraphics[width=0.22\columnwidth]{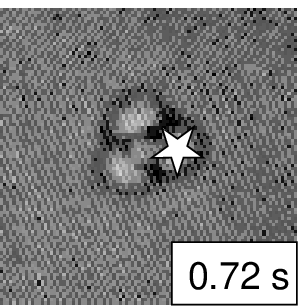}&
\includegraphics[width=0.22\columnwidth]{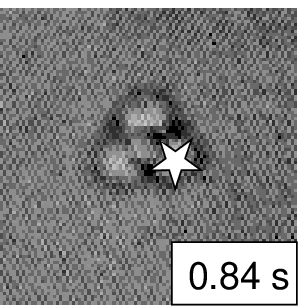}\\
\includegraphics[width=0.22\columnwidth]{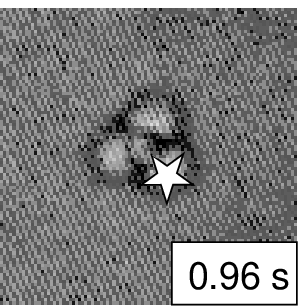}&
\includegraphics[width=0.22\columnwidth]{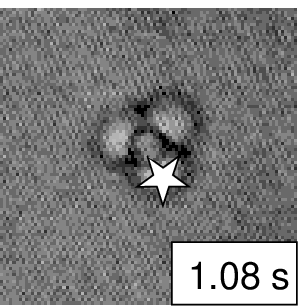}&
\includegraphics[width=0.22\columnwidth]{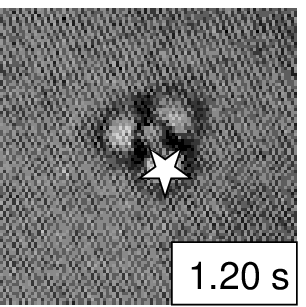}&
\includegraphics[width=0.22\columnwidth]{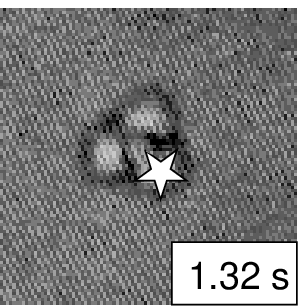}\\
\includegraphics[width=0.22\columnwidth]{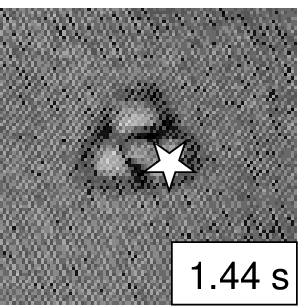}&
\includegraphics[width=0.22\columnwidth]{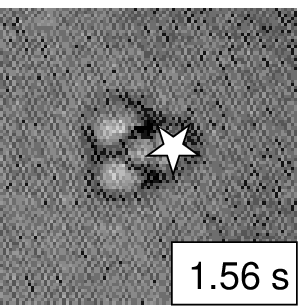}&
\includegraphics[width=0.22\columnwidth]{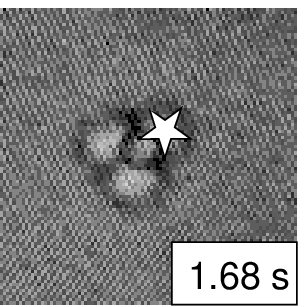}&
\includegraphics[width=0.22\columnwidth]{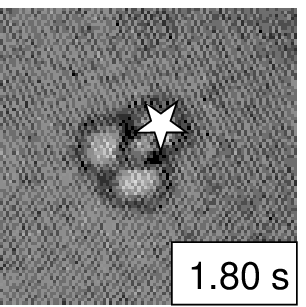}\\
\includegraphics[width=0.22\columnwidth]{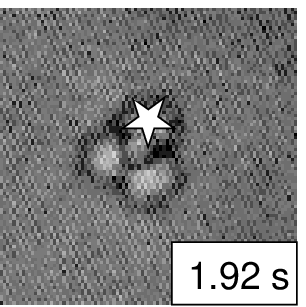}&
\includegraphics[width=0.22\columnwidth]{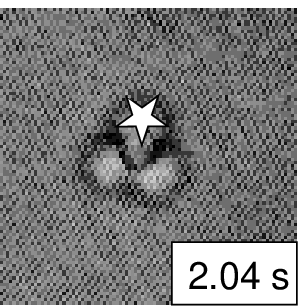}&
\includegraphics[width=0.22\columnwidth]{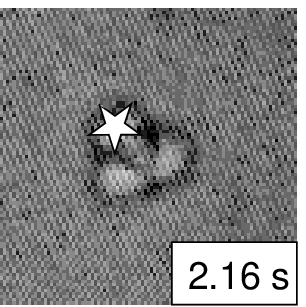}&
\includegraphics[width=0.22\columnwidth]{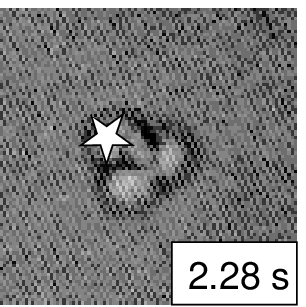}
\end{tabular}
\caption{Reversal of rotation of particles trapped in a helical beam.
The helicity of the beam is reversed at time 1.20\,s by removing a Dove
prism from the beam path, causing the rotation of the clump of weakly
absorbing polystyrene spheres to be reversed.}
\end{figure}

This method of detection has serious limitations, as particles move out of
clear focus when they are trapped and so their images are not well defined
on video.  In the second set of experiments the experimental set-up was
refined in two ways.  Firstly, a quarter-wave plate was introduced into
the beam path directly before the $100\times$ microscope objective
(NA$ = 1.3$) which, as before, focused the doughnut beam to a waist
size of approximately 2\,{\textmu}m in diameter.  The purpose of this
experiment was to allow a precise measurement of the rotation speed and
also to study the influence of the added circular polarisation of the beam
by which the theoretical prediction of the total transfer of angular
momentum could be verified.

A second refinement was made to allow for precise measurement of the
rotation frequency.  The rotation was this time measured using a
photodetector positioned off centre to detect a portion of the light
reflected from the rotating particle.  The particles are irregularly
shaped and the protruding parts of the particles reflect a ``flash'' of
light onto the displaced detector.  The signal of intensity fluctuations
over time obtained from the photodetector is Fourier transformed yielding
the rotation frequency of the particle.  The modified experimental set-up
for the above experiments is shown in Figure 3.

The sequence of the experiment is as follows. The LG$_{03}$ beam enters
the microscope.  It passes through a $\lambda/4$ plate and its focus is
adjusted to be slightly below the focal point of the microscope.  The
polarisation state is checked after the objective lens by an analyser and
the power of the beam is monitored for each position of the plate (left
and right-circularly polarised light and linearly polarised light).  It
was established that the power output varied less than 1\% between
different polarisations.  Care was taken to position the $\lambda/4$ plate
in such a way as not to cause any deflection of the beam when the plate was
rotated.  The sample of absorbing CuO particles of sizes up to
20\,{\textmu}m in kerosene was placed between a glass microscope slide
and coverslip.  A particle was trapped into the linearly polarised helical
doughnut and the $\lambda/4$ plate rotated firstly to right circular
polarisation, back to linear and then to left circular polarisation.
The signal was sampled at 20\,Hz for a period of 100\,s.  The sequence
was repeated and the reflected light monitored on the photodetector.
An example of the spectrum resulting from this experiment is shown in
Figure 5. It can be seen from this figure that rotation frequency increases
when the helicity of the electric field vector has the same direction as the
helicity of the Poynting vector and decreases when these directions are
opposite to each other.

\begin{figure}[ht]
\includegraphics[width=\columnwidth]{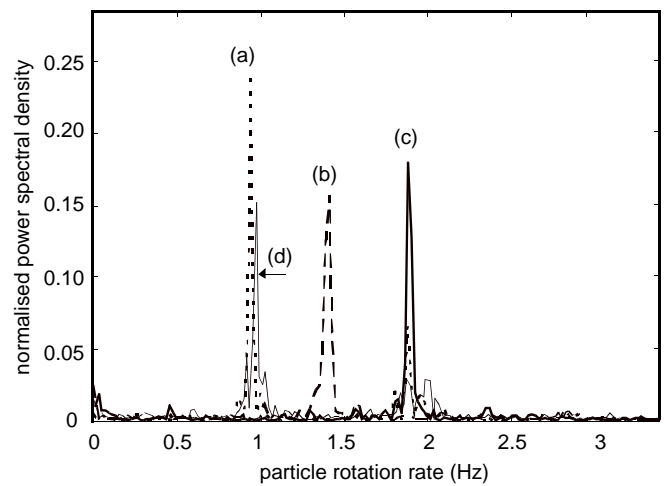}
\caption{Rotation speeds for absorbing particles trapped by a helical
doughnut beam when the polarisation state of the trapping beam is varied.
Power spectra obtained by Fourier analysis of the photodiode signal
obtained from light scattered off rotating CuO particles.  The peaks
at (a), (b), and (c) represent the  measured rates of rotation of CuO
particles trapped in a left-circularly polarised right-helical, plane
polarised right-helical, and right-circularly polarised right-helical
doughnut beam respectively. The peak at (d)  is the  4th set of data,
taken with left-circularly polarised right-helical light, in order to
verify that this rotation rate did not vary during the experiment. The
particle's  rotation rates are approximately in the ratio 2:3:4.}
\end{figure}

Using an LG$_{03}$ doughnut the theoretical prediction is that the
frequencies should scale as 2:3:4 when going from opposite helicity
Poynting vector and spin to linearly polarised helical wave front
through to the same helicities.  This was verified experimentally
(see Figure 5).

Recently a somewhat similar experiment has been performed by
Simpson et al.~\cite{ref23} using a charge one Gauss--Laguerre laser mode
produced by operating a laser in a Hermite--Gaussian mode and converting
the output into the corresponding Gauss--Laguerre mode using a cylindrical
lens mode converter.  A quarter-wave plate was used to change polarisation.
The authors showed that spin angular momentum of $\pm\hbar$ per photon
associated with circularly polarised light can add to, or subtract from,
the orbital angular momentum and observed the mutual cancellation of the
spin and orbital angular momentum.  Their results confirm the results
obtained by our group and show that the LG$_{01}$ mode of charge 1 has a
well-defined orbital angular momentum corresponding to $\hbar$ per photon.

In our most recent experiments we have shown that an absorbing particle can
in fact be trapped by a Gaussian beam.  Two-dimensional trapping can be
achieved before the waist of a tightly focused Gaussian beam, where the
spot size is rapidly changing, against a surface such as a microscope
slide, using radiation pressure.  The particle was first trapped using
linearly polarised Gaussian beam and subsequently by turning a $\lambda/4$
plate a circularly polarised Gaussian beam was used to trap the particle
resulting in a rotation of this particle.  When the $\lambda/4$ plate was
rotated to produce circularly polarised light of the opposite sense, the
particle reversed the direction of rotation.  Successive frames from a
video demonstrating trapping are shown in Figure 6.

\begin{figure}[hb]
\begin{tabular}{cccc}
(a) &
\includegraphics[width=0.3\columnwidth]{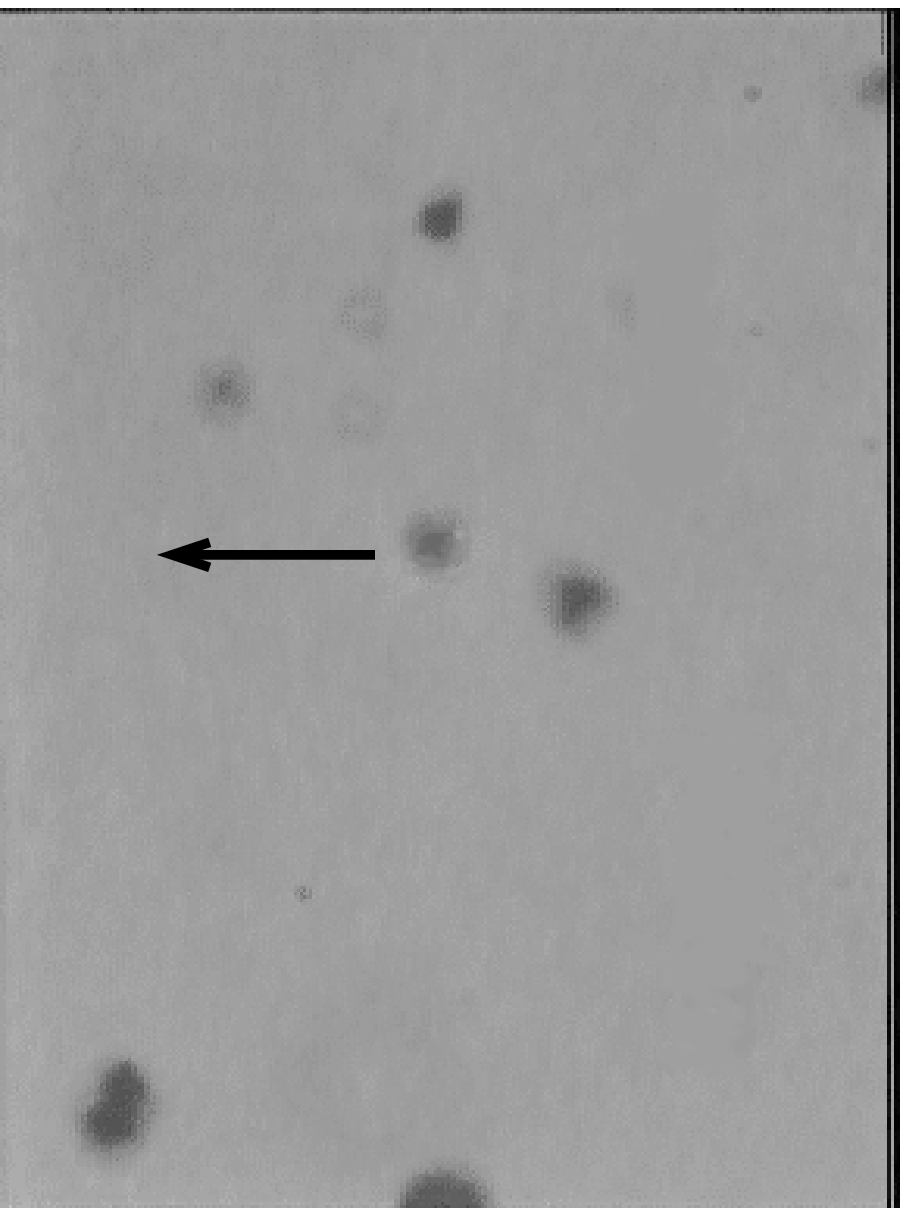}&
\includegraphics[width=0.3\columnwidth]{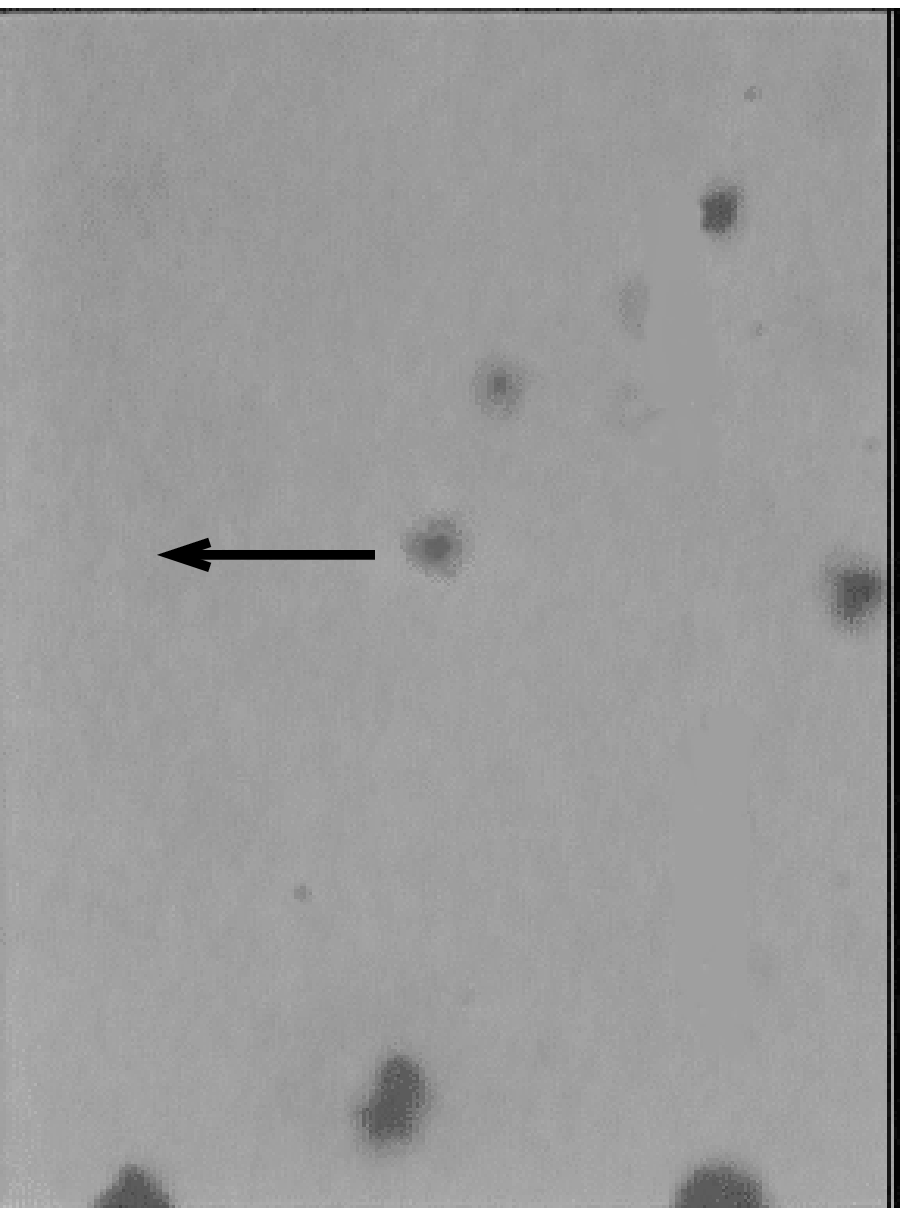}&
\includegraphics[width=0.3\columnwidth]{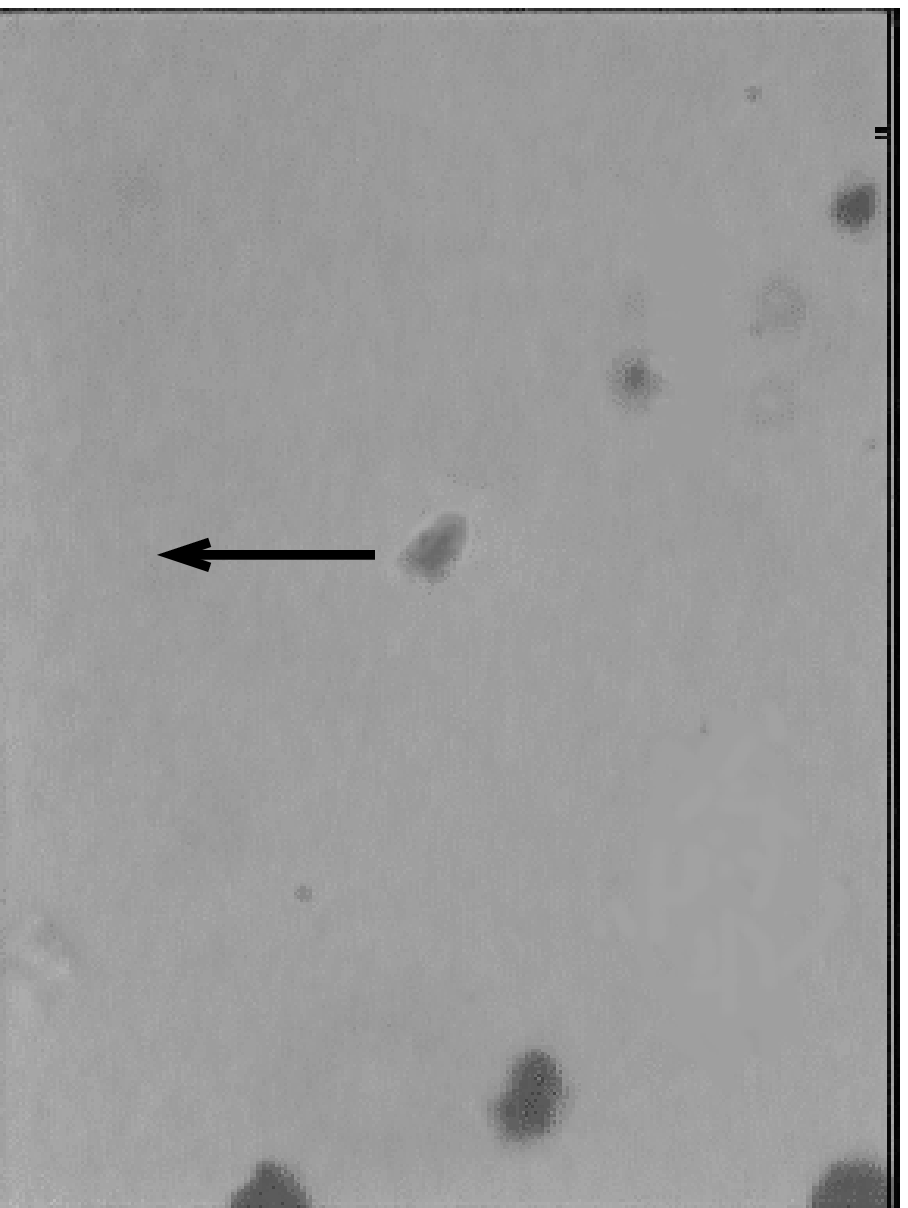}\\
(b) &
\includegraphics[width=0.3\columnwidth]{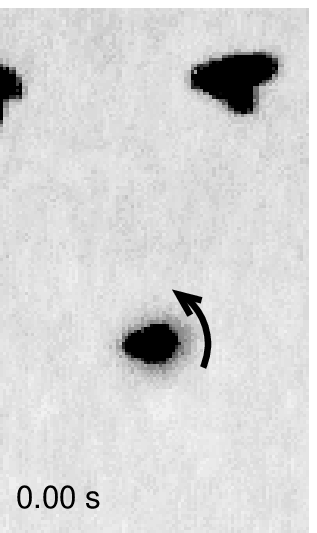}&
\includegraphics[width=0.3\columnwidth]{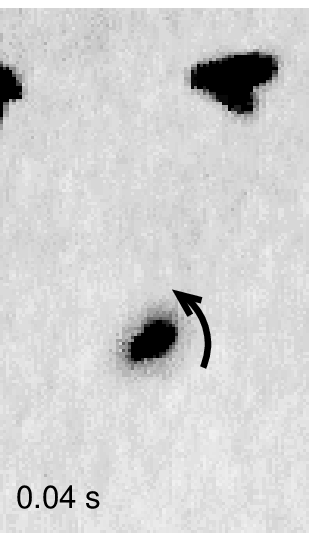}&
\includegraphics[width=0.3\columnwidth]{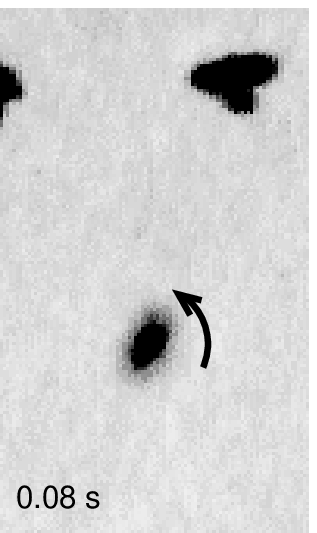}
\end{tabular}
\caption{CuO particle trapped in a Gaussian beam.
Frames from a video recording CuO particles being trapped, moved and
rotated using a Gaussian beam.
(a) shows motion with respect to the surrounding particles and (b) shows
rotation produced by circular polarisation of the beam.}
\end{figure}

\section{Theory of trapping of absorbing particles}

The physical basis of the trapping of absorbing particles can be
investigated in terms of the interaction between radiation and matter.
For highly absorbing particles, effects such as induced polarisation
within the particle, refraction and Doppler shifts which are important
for transparent particles and atoms can be neglected.  The properties of
the trap depend mainly on the properties of the trapping beam, which will
be assumed to be paraxial.  While a non-paraxial beam is necessary for
laser trapping, the use of the paraxial approximation does not seem to
introduce excessive error.  Errors due to aberration within the optical
system are expected to be greater.

The theory will be developed in terms of the two most common types of
beams used for trapping particles, namely TEM$_{00}$ Gaussian beams
[denoted (G)] and LG$_{pl}$ Laguerre--Gaussian ``doughnut'' beams
[denoted (LG)]
described by a radial mode index $p$ and an azimuthal mode index $l$.

\subsection{The transfer of linear momentum and orbital angular momentum}

We begin with consideration of the  transfer of linear momentum to an
absorbing infinitesimal element.  This microscopic behaviour can then be
extended to the particle as a whole.  In the course of this, it can be seen
that the transfer of orbital angular momentum from a helical beam to a
particle is a simple radiation pressure process involving the transfer of
linear momentum to different portions of the particle.  Due to the spatial
structure of the helical beam, this results in a transfer of angular momentum. 

\subsection{Microscopic transfer of momentum}

Due to the cylindrical symmetry of the beam, it is convenient to use a
cylindrical coordinate system, with radial, azimuthal and axial coordinates
$r$, $\phi$, and $z$, and corresponding
unit vectors $\hat{\mathbf{r}}$, $\hat{\phi}$, and $\hat{\mathbf{z}}$.
The linear momentum flux of a laser beam is given by the time-averaged
Poynting vector $\mathbf{S}$ (in cylindrical coordinates)~\cite{ref48}:
\begin{equation}
\mathbf{S} = \frac{c\epsilon}{2} E_0^2
\left( \frac{zr}{z_r^2+z^2} \hat{\mathbf{r}} + \hat{\mathbf{z}} \right)
\;\;\;\; \mathrm{(G)}
\end{equation}
\begin{equation}
\mathbf{S} = \frac{c\epsilon}{2} E_0^2
\left( \frac{zr}{z_r^2+z^2} \hat{\mathbf{r}}
+ \frac{l}{kr} \hat{\phi} + \hat{\mathbf{z}} \right)
\;\;\;\; \mathrm{(LG)}
\end{equation}
where $z_r$ is the Rayleigh range, $k$ is the wavenumber of the beam, and
$E_0$ is the amplitude of the beam, given by~\cite{ref49}

\begin{figure*}[p]
\includegraphics[width=\textwidth]{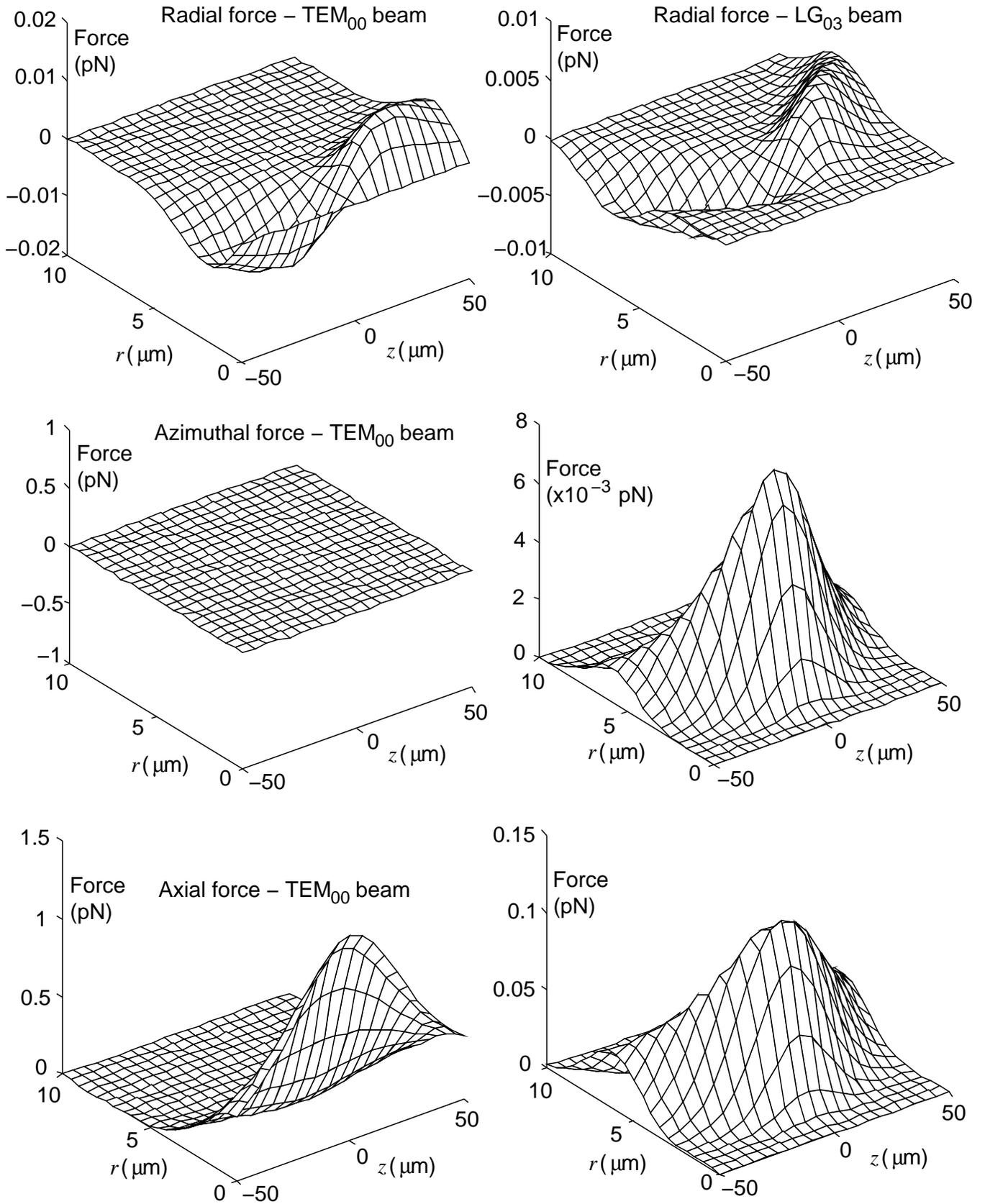}
\caption{Spatial dependence of optical force on an absorbing particle.
The radial and axial variation of the optical force is shown for both a
TEM$_{00}$ Gaussian beam and an LG$_{03}$ Laguerre--Gaussian beam.  Both
beams have the same power (1\,mW), spot size (2\,{\textmu}m) and wavenumber
(free space wavelength 632.8\,nm).  The particle has a circular cross-section
of radius 1\,{\textmu}m.  Due to the cylindrical symmetry, there is no
azimuthal variation of the force.  The beam is propagating in the $+z$
direction, with the beam waist at $z = 0$.}
\end{figure*}

\newpage

\begin{equation}
E_0 = \sqrt{\frac{2}{c\epsilon}} \sqrt{\frac{2P}{\pi}}
\frac{1}{w(z)} \exp\left( \frac{-r^2}{w^2(z)} \right)
\;\;\;\; \mathrm{(G)}
\end{equation}
\begin{eqnarray}
E_0 & = & \sqrt{\frac{2}{c\epsilon}} \sqrt{\frac{2p!P}{\pi(p+l)!}}
\left(\frac{r\sqrt{2}}{w(z)}\right)^l
L_p^l\left( \frac{2r^2}{w^2{z}} \right) \times \nonumber \\ & &
\frac{1}{w(z)} \exp\left( \frac{-r^2}{w^2(z)} \right)
\;\;\;\; \mathrm{(LG)}
\end{eqnarray}
where $P$ is the beam power, $w(z)$ is the beam width, $L_p^l(x)$ is the
generalised Laguerre polynomial, $c$ is the speed of light and $\epsilon$
is the permittivity.  The Rayleigh range $z_r$ and the beam width $w(z)$
are related to each other and the beam waist spot size $w_0$ by:
\begin{equation}
w^2(z) = \frac{2(z_r^2+z^2)}{kz_r}
\end{equation}
\begin{equation}
w(z) = w_0 \sqrt{1+z^2/z_r^2}
\end{equation}
\begin{equation}
z_r = kw_0^2/2
\end{equation}
For most cases of interest involving doughnut beams, the radial mode index
$p = 0$ and the Laguerre polynomial $L_p^l = 1$.  If an area element
$\mathrm{dA}$ is highly absorbing, the rate of momentum transfer to it
will be given by the Poynting vector $\mathbf{S}$:
\begin{equation}
\mathbf{F} = \frac{1}{c}\mathbf{S}\cdot(-\mathbf{dA})
\frac{\mathbf{S}}{|\mathbf{S}|}
\end{equation}
The transfer of momentum from the beam to an absorbing particle is
therefore straightforward compared to other cases, such as transparent
particles and atoms.  For transparent particles, refraction and induced
polarisation must be taken into account.  For an atom, the frequency
dependence of the absorption and spontaneous emission must be considered,
while for an absorbing particle, the absorption can be assumed to be
independent of frequency, and inter-atomic collision rates within the
particle can be assumed to be high enough to cause de-excitation without
re-emission.

\subsection{Macroscopic effects}

As particles typically are sufficiently large that the Poynting vector of
the laser beam is not constant over their absorbing surfaces, it is
necessary to integrate the Poynting vector over the surface.  This requires
knowledge of the geometry of the particle and the beam.  This calculation
can be readily performed numerically by dividing the particle into a number
of small area elements dA over which the Poynting vector is approximately
constant, and integrating equation (10).  Where the product
$\mathbf{S}\cdot(-\mathbf{dA}) < 0$, the particle will not be illuminated,
so such regions can be neglected.  A useful and simple approximation is to
simply represent the particle as a disk with the same cross-section
(essentially applying the paraxial approximation to the illumination of
the particle).  The resultant force on a particle is shown in Figure~7.

Equilibrium points in these force fields exist only where the force is zero,
and a particle can be trapped only at a stable equilibrium point.  Thus, it
can be readily seen that a particle cannot be axially trapped without an
external force (such as gravity, viscous drag due to convection, or a
reaction force due to the particle resting on the bottom of the trapping
cell) acting on it.  The particle can be trapped radially in the portion
of the beam which is converging towards the beam waist.  The particle
cannot be trapped after the beam waist, as the convergence of the beam
required for trapping no longer exists.  The absorbing particle trap can
therefore be considered to be a two-dimensional trap.  Also, it should be
noted that in the case of Laguerre--Gaussian beams, angular momentum is
transferred to the particle as a whole, although microscopically only
linear momentum is seen to be present.

The relative efficiencies of trapping absorbing particles using Gaussian
beams and doughnut beams can be measured in terms of the ratio of the
radial force to the axial force.  As can be seen from Figure 8, the
doughnut beam trap has smaller axial forces for trapped particles.

\begin{figure}[hb]
\includegraphics[width=\columnwidth]{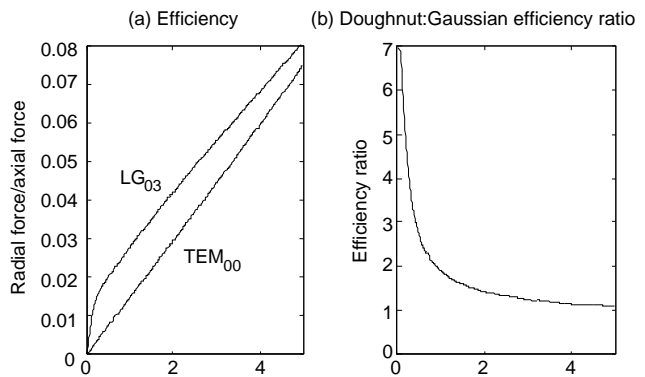}
\caption{Efficiencies of trapping beams.
The efficiency of the trap can be measured by the ratio of the radial force
to the axial force.  These ratios are calculated here for beams of waist size
2\,{\textmu}m and power of 1\,mW.  The particle has a radius of 1\,{\textmu}m
and is in a plane 30\,{\textmu}m before the waist.}
\end{figure}

\begin{figure*}[p]
\includegraphics[width=\textwidth]{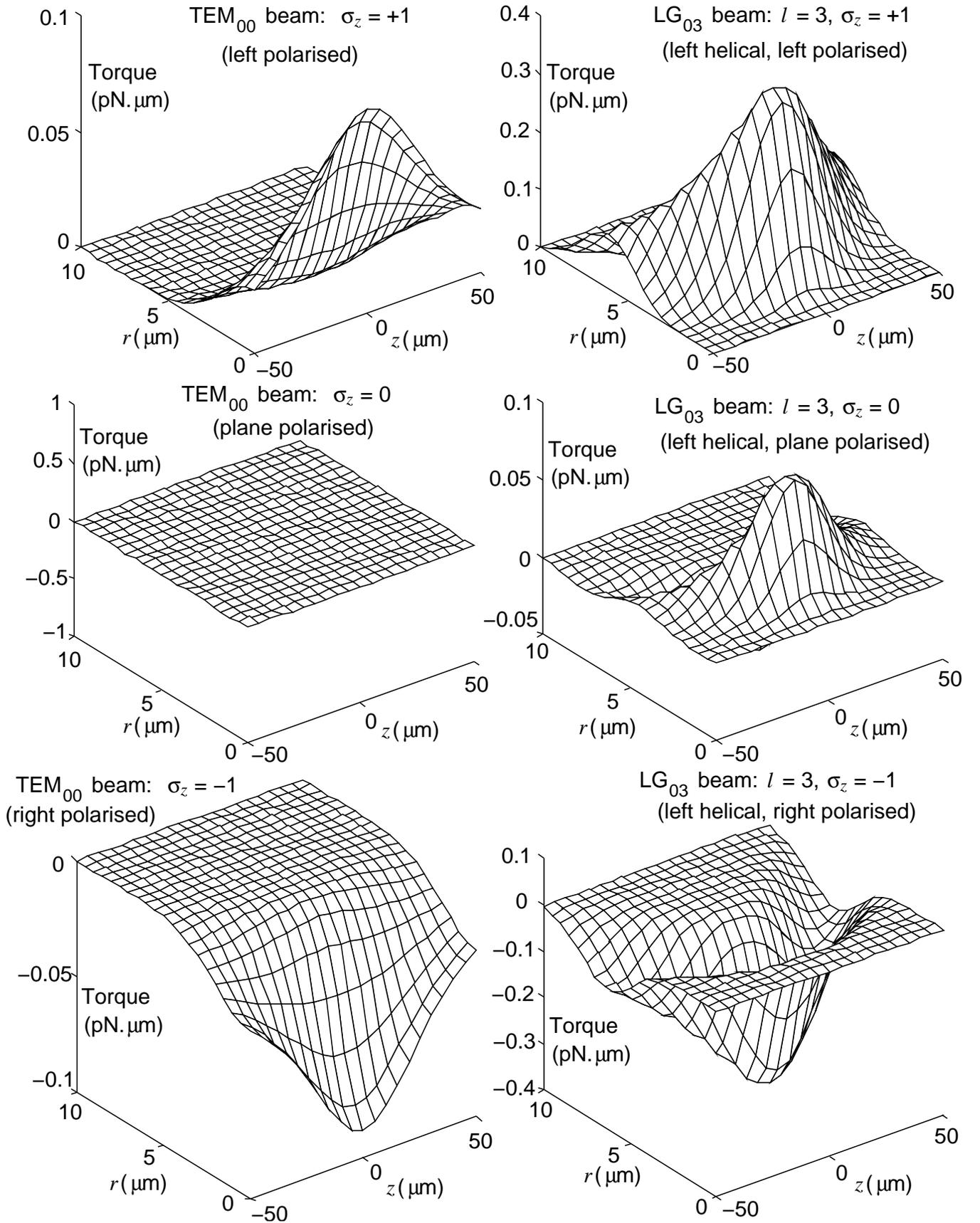}
\caption{Spatial dependence of torque parallel to beam axis on absorbing
particle. The torque experienced by a 1\,{\textmu}m radius particle through
absorption of a combination of spin and orbital angular momentum is shown
for both a Gaussian TEM$_{00}$ beam and a left-helical LG$_{03}$ doughnut
beam.  In all cases, the beams have a waist size of 2\,{\textmu}m, a power
of 1\,mW, and a free space wavelength of 632.8\,nm.  Other components of
the torque will be much smaller.}
\end{figure*}

As the force acting on a particle at any position within the beam can be
calculated, the overall properties of the trap, such as the trapping force,
or the motion of trapped particles can be determined.  If particles are
confined to a two-dimensional plane in which they are radially trapped,
particles trapped in Gaussian beams are pushed into the centre of the trap,
while particles trapped in Laguerre--Gaussian beams also tend to orbit about
the beam axis due to the azimuthal component of the force.  This orbital
motion can lead to instability in the trap, as a particle undergoing such
motion will only remain within the trap if the radial force can provide a
centripetal force sufficient to force the path to become circular.  If the
viscosity of the surrounding medium is too low, the terminal orbit speed
will be high enough so that the radial force will be unable to provide the
necessary centripetal acceleration and the particle will escape from the
trap.  Particles trapped in fluids such as water or kerosene, however, have
very low terminal speeds and can be trapped.

\section{Transfer of Spin Angular Momentum}

As seen above, any orbital angular momentum present in a helical laser
beam (for example, a Laguerre--Gaussian beam) is transferred to a particle
by the same mechanisms as linear momentum.  All laser beams, however, can
carry angular momentum if the beam is circularly polarised, with each photon
having angular momentum of magnitude $\hbar$.  The rate of absorption of
angular momentum by a small section of the particle is given in terms of
the Poynting vector $\mathbf{S}$ and the wavenumber $k$ by
\begin{equation}
\tau = \frac{\sigma_z}{k}\mathbf{S}\cdot(-\mathbf{dA})
\frac{\mathbf{S}}{|\mathbf{S}|}
\end{equation}
where $\sigma_z = \pm 1$ for left- and right-circular polarisation and
$\sigma_z = 0$ for plane polarisation.

The total torque on the particle due to polarisation can then be found by
integrating equation (11) over the absorbing surface of the particle (see
Figure 9).  The rotational behaviour of particles trapped in a Gaussian
beam is straightforward.  However, that of  a particle illuminated by a
Laguerre--Gaussian beam is more complex, due to the presence of the
orbital angular momentum.  In practice a particle is trapped on the
beam axis.  The total torque due to polarisation on a particle of
radius $r$ trapped on the beam axis is given by
\begin{equation}
\tau_\mathrm{p} = \frac{\sigma_z P}{\omega}
\left\{ 1- \exp\left( \frac{-2r^2}{w^2(z)} \right) \right\}
\;\;\;\; \mathrm{(G)}
\end{equation}
\begin{equation}
\tau_\mathrm{p} = \frac{\sigma_z P}{\omega}
\left\{ 1- \exp\left( \frac{-2r^2}{w^2(z)} \right)
\sum_{k=0}^l \frac{1}{k!} \left( \frac{2r^2}{w^2(z)} \right)^k \right\}
\;\;\; \mathrm{(LG)}
\end{equation}
where $\omega$ is the angular frequency of the light.  The total torque
can be found by combining this with the torque due to the orbital angular
momentum carried by the helicity of the beam
\begin{equation}
\tau_\mathrm{o} = 0
\;\;\;\; \mathrm{(G)}
\end{equation}
\begin{equation}
\tau_\mathrm{o} = \frac{lP}{\omega}
\left\{ 1- \exp\left( \frac{-2r^2}{w^2(z)} \right)
\sum_{k=0}^l \frac{1}{k!} \left( \frac{2r^2}{w^2(z)} \right)^k \right\}
\;\;\; \mathrm{(LG)}
\end{equation}
The angular velocity $\Omega$ in a viscous medium will depend on the drag
torque, which for a smooth spherical particle of radius $r$ in a medium
of viscosity $\eta$ is given by~\cite{ref50}
\begin{equation}
\tau_\mathrm{d} = -8\pi\eta r^3\Omega								(16)
\end{equation}
and the optical torque.  The resultant spin rates for particles trapped
on the beam axis are
\begin{equation}
\Omega = \frac{\sigma_z P}{8\pi\eta r^3\omega}
\left\{ 1- \exp\left( \frac{-2r^2}{w^2(z)} \right) \right\}
\;\;\;\; \mathrm{(G)}
\end{equation}
\begin{eqnarray}
\Omega & = & \frac{(\sigma_z +l)P}{8\pi\eta r^3\omega}
\left\{ 1- \exp\left( \frac{-2r^2}{w^2(z)} \right)
\sum_{k=0}^l \frac{1}{k!} \left( \frac{2r^2}{w^2(z)} \right)^k \right\}
\nonumber \\ & &
\hspace{5cm} \mathrm{(LG)}
\end{eqnarray}

\begin{figure}[b]
\includegraphics[width=\columnwidth]{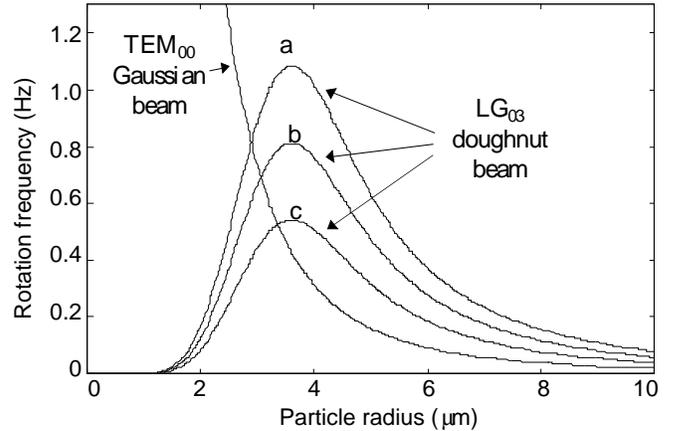}
\caption{Spin rates of trapped absorbing particles trapped in kerosene.
A (two-dimensionally) trapped particle will experience a torque dependent on
its size and the beam power.  This torque will depend on the position along
the beam axis where the particle is trapped and on the direction of circular
polarisation of the beam.
The beams here are of width 2\,{\textmu}m and power 1\,mW.  The three cases
for particles trapped by a LG$_{03}$ doughnut beam are shown by: (a)
circular polarisation and helicity in the same direction, (b) plane
polarisation, and (c) circular polarisation and helicity
in opposite directions.}
\end{figure}

The rotation rate for particles trapped in a Gaussian beam rises as the
particles become small.  When the spin rate becomes large, equation (16)
will cease to be applicable.  For particles trapped in the doughnut beam,
the rotation rate is maximum for a particle large enough to absorb most of
the beam, but small enough to keep the drag surface area small.

In a typical experiment, ceramic particles with radii of about 2\,{\textmu}m,
suspended in kerosene, were trapped in a beam of 4\,mW power and a width of
1.75\,{\textmu}m.  Rotation rates of 1--2\,Hz were observed (see Figure 5),
consistent with the predictions of equation (18) and Figure 10.

\section{Absorption of Energy}

The energy density in a focussed beam of even modest power can be very large.
For example, in a 1\,mW beam focussed to a 1\,{\textmu}m waist, the
irradiance will exceed $10^9$\,W/m$^2$.  The transparent objects
conventionally trapped by optical tweezers evidently absorb very
little, partly because they are very thin, and they are surrounded by a
conducting medium so experience only modest rises in temperature.  Indeed,
it has been shown that bacteria will survive and even reproduce while
trapped.

The situation for strongly absorbing particles is very different.  A typical
1\,{\textmu}m radius particle absorbing 1\,mW will, in the absence of any
losses, experience a temperature rise of order $10^8$\,K/s.  Clearly,
the temperature of a trapped particle will rise rapidly and it is necessary
to consider what losses will limit the rise.

Since the surface area is small, even neglecting local absorption in
the surrounding medium, the amount of heat which can be radiated is
limited.  Stefan's Law shows that a particle of 1\,{\textmu}m radius needs
to have a surface temperature in excess of 1000\,K before the power radiated
exceeds 1\,{\textmu}W.  The small surface area also severely limits the
rate at which heat can be exchanged with the surrounding fluid.

If we consider a spherical particle in equilibrium losing the heat
absorbed from the beam by isotropic conduction into a surrounding
medium of thermal conductivity $k$, the radial variation of the temperature
$T$ in the surrounding medium can easily be shown to be given by
\begin{equation}
T = \frac{P_\mathrm{a}}{4\pi rk} + T_0
\end{equation}
where $P_\mathrm{a}$ is the total power absorbed by the particle from
the beam and $T_0$ is the ambient temperature.  The absorbed power for
a particle with absorption coefficient $\alpha$ trapped on the beam axis of a
TEM$_{00}$ Gaussian or LG$_{0l}$ Laguerre-Gaussian beam of power $P$ is
given by
\begin{equation}
P_\mathrm{a} = \alpha P
\left\{ 1- \exp\left( \frac{-2r^2}{w^2(z)} \right) \right\}
\;\;\;\; \mathrm{(G)}
\end{equation}
\begin{equation}
P_\mathrm{a} = \alpha P
\left\{ 1- \exp\left( \frac{-2r^2}{w^2(z)} \right)
\sum_{k=0}^l \frac{1}{k!} \left( \frac{2r^2}{w^2(z)} \right)^k \right\}
\;\;\;\; \mathrm{(LG)}
\end{equation}

Calculated equilibrium temperatures reached by typical particles trapped in a
1\,mW beam are shown in Figure 11.

\begin{figure}[ht]
\includegraphics[width=\columnwidth]{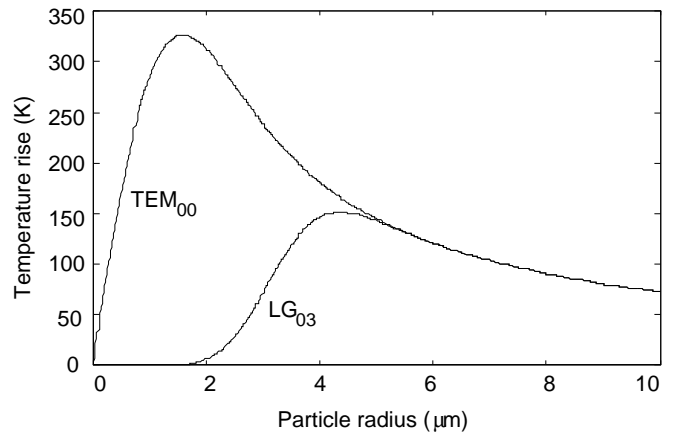}
\caption{Temperature rise of trapped particles.
The temperature rise due to power absorbed from a laser beam is shown as a
function of particle size.  The trapping beams have a power of 1\,mW and a
width of 2\,{\textmu}m where the particle is trapped.  Small particles
absorb little energy from the beam while very large particles lose
significantly more heat through conduction due to their large surface area.}
\end{figure}

Quite high temperatures are predicted, especially for particle sizes
on the order of the spot size.

Convection is another possible means of heat loss.  The importance of
convection can be found in terms of the mean Nusselt number $\bar{Nu}$,
which, at low flow rates, is given in terms of the Prandtl number $Pr$
(the ratio between the momentum and thermal diffusivities) and the Grashof
number $Gr$ (the ratio between buoyant and viscous forces) by~\cite{ref51}
\begin{equation}
\bar{Nu} = 2 + Gr + Gr^2(0.139-0.419Pr+1.1902Pr^2)
\end{equation}
The Grashof number is given by~\cite{ref52}
\begin{equation}
Gr = g(1-\rho_T/\rho_\infty)D^3/\nu^2
\end{equation}
and the Prandtl number by
\begin{equation}
Pr = c_p\eta/k
\end{equation}
where $c_p$ is the specific heat of the fluid, $\eta$ is the dynamic
viscosity and $\nu$ is the kinematic viscosity.

For a temperature difference of 100\,K and an ambient temperature of 300\,K,
and a particle diameter of $D = 3$\,{\textmu}m (at which size high
temperatures are reached for a typical beam), $\bar{Nu}=2+3\times 10^{-4}$
for water and $\bar{Nu} = 2 + 7\times 10^{-5}$ for kerosene, values which
are very close to $\bar{Nu} = 2$ for pure conduction.  The convective
contribution to the cooling is therefore negligible.

We are therefore driven to the conclusion that even in the presence of a
liquid medium, an absorbing particle will experience a large rise in
temperature, probably limited by thermal conduction.  Note that due to
the $1/r$ dependence of the temperature rise in that case, the volume
where elevated temperatures will be found is quite small which means in
turn that the thermal relaxation time will be short, so that the temperature
rises will be difficult to measure.

However, we do have evidence of such large temperature rises.  In
experiments where we trapped small particles of photocopy toner, we
observed that initially jagged fragments almost instantly became smooth
when irradiated by a beam with a power of a few mW.  According to the
manufacturer's specifications, the fusing point of the toner is
185{\textdegree}C.  These results are at first rather surprising as it
might have been expected that boiling of the liquid would enhance cooling
or disrupt the trapping process.  Sometimes, with beam powers well in excess
of that required to melt toner, a bubble is generated but it is not clear if
this is vapour or merely absorbed gases being evolved from the particle.
The boiling process in general is a somewhat mysterious one and it is
widely accepted that bubble formation requires some seed to overcome
surface tension forces which scale inversely as the radius.  Therefore
it cannot be taken for granted that normally observed behaviour will
scale to very small dimensions.  If so, further study of superheated
particles may help shed new light on the fundamentals of the boiling
process.

\section{Interaction with the Medium}

Interactions with the surrounding transparent medium include the heating
of the medium through conduction from the absorbing particle, and the
generation of fluid motions through stirring and convection currents.

Temperature changes in the medium will be localised to within 100\,{\textmu}m
of the particle but could affect properties such as the refractive index and
viscosity.  It is also possible that thermophoretic effects may occur in the
strong temperature gradients formed.

Stirring and convection effects have both been observed when CuO particles
were trapped, using a doughnut beam, in water to which a little detergent
had been added.  This greatly reduced the tendency of particles to stick
to the slide and generally freed up their motion so that Brownian motion
was more evident.  When a particle was trapped, others in its vicinity
began to move toward it, presumably as a result of a convection currents,
and began to circle around the trapped particle in the same direction as
its rotation, presumably carried around by a flow generated by the rotation
of the trapped particle.  The fluid velocities of the order of
10\,{\textmu}m/s are too small to significantly affect heat transport.

\section{Conclusion}

It has been shown that the converging section of a focussed laser beam
can radially trap absorbing particles against a supporting substrate
using a Gaussian or doughnut beam.  Doughnut beams with phase
singularities carry angular momentum which is also transferred to the
trapped particle causing it to rotate.  Angular momentum resulting from
circular polarisation of the light can also be transferred.  Energy
absorbed from the beam can lead to rapid heating and high equilibrium
temperatures, even well in excess of the normal boiling point of the
surrounding medium.



\end{document}